# Dominant role of charge ordering on high harmonic generation in $Pr_{0.6}Ca_{0.4}MnO_3$

A. Nakano [1*], K. Uchida[1], Y. Tomioka[2], M. Takaya[3], Y. Okimoto[4] and K. Tanaka[1,5*]

[1] *Department of Physics, Graduate School of Science, Kyoto University, Kyoto, Kyoto 606-8502, Japan*

[2] *National Institute of Advanced Industrial Science and Technology (AIST), Tsukuba, 305-8565, Japan.*

[3] *Department of Geology and Mineralogy, Graduate School of Science, Kyoto University, Kyoto, Kyoto 606-8502, Japan*

[4] *Department of Chemistry, Tokyo Institute of Technology, Meguro, Tokyo 152-8551, Japan*

[5] *Institute for Integrated Cell-Material Sciences, Kyoto University, Sakyo-ku, Kyoto 606-8501, Japan*

*\*e-mail: nakano.aiko.75n@st.kyoto-u.ac.jp, kochan@scphys.kyoto-u.ac.jp*

## Abstract

High-harmonic generation (HHG) is a typical high-order nonlinear optical phenomenon and can be used to probe electronic structures of solids. Here, we investigate the temperature dependence of HHG from $Pr_{0.6}Ca_{0.4}MnO_3$ in the range of 7 K to 294 K including the charge ordering (CO) transition and magnetic transition temperatures. The high-harmonic intensity remains almost constant in the high-temperature charge-disordered phase. However, as the temperature is lowered, it starts to gradually increase near the CO transition temperature where an optical gap related to the CO phase appears. The anomalous gap energy dependence resembles the one recently reported in a Mott insulator. We attribute the HHG suppression at high temperatures to the destructive interference among high-harmonic emissions from thermally activated multiple CO configurations. Our results suggest that HHG is a promising tool for probing the fluctuation of local order in strongly correlated systems.

High-order harmonic generation (HHG) is typical of extreme nonlinear optical phenomena resulting from the interaction of materials with intense light and was first reported to occur in atomic gases [1]. The origin of HHG is attributed to sub-laser-cycle carrier dynamics driven by the laser's electric field. In crystalline solids, the high harmonics originate from the motion of Bloch electrons in a periodic potential, meaning that HHG measurements allow us to probe the properties of the material [3-10]. For example, the momentum-dependent bandgap of ZnO along the $\Gamma - M$ direction was reconstructed by using high harmonics from a two-color laser pulse [6].

The properties of HHG in typical semiconductors can be understood through a semiclassical three-step model similar to that in atomic gasses [2], consisting of Zener-tunneling, acceleration, and recombination of Bloch electrons [11-14]. On the other hand, this simple picture does not necessarily hold in strongly correlated systems (SCSs) because the strong coupling of charge with multi degrees of freedom modifies the charge dynamics driven by the laser field, which emits high harmonics. Although many theoretical studies have predicted unique aspects of HHG in SCSs [15-22], only a few experiments have been performed [23-25].

The recent HHG measurement of a Mott insulator $Ca_2RuO_4$ has demonstrated that HHG in SCSs offers unique nonlinear optical phenomena peculiar to SCSs and has the potential to probe many-body states in SCSs [25]. In the measurement, the intensity of the harmonics increased as the gap energy increased with decreasing temperature. This is contrary to what one would intuitively expect from the three-step model, where the tunneling probability should be smaller as the gap energy increases. This unusual gap energy dependence was qualitatively reproduced in the simulation of a single-band Hubbard model at finite temperatures [22]. The authors in Ref. 22 proposed a scenario that the temperature-dependent spin background affects the charge dynamics through spin-charge coupling as follows: In the ground state, the spins are arranged antiferromagnetically. That is, if one site has an electron with spin up, the next neighboring site should have an electron with spin down at the ground state. As the temperature of the system gets higher, more spin configurations different from the antiferromagnetically ordered ground state are thermally activated. Considering that there is spin-charge coupling, the emitted light resulting from the charge motion would have different phases for different spin configurations. Therefore, the destructive interference of high-harmonic emissions from thermally disturbed multiple spin configurations may suppress the high-harmonic intensity at finite temperatures.

The above scenario is not restricted to the materials with spin-charge coupling but could apply more generally to materials where other degrees of freedom are coupled to charge dynamics. A material with a charge-ordered (CO) phase is an attractive candidate to verify the universality of the scenario. In its charge-ordered ground state, charges are periodically localized in space. As in the case of the antiferromagnetic (AFM) phase, the CO configuration is thermally disturbed and deviates from the ground state at finite temperatures. If the charge ordering is strongly coupled with the charge dynamics, this system may exhibit HHG properties similar to those reported for the Mott insulator [22].

In this letter, we report the temperature-dependent properties of HHG in manganese oxide with several electronic phases including the charge-ordered (CO) phase. Our purpose is to clarify the effect of the couplings of the charge dynamics with the multi degrees of freedom on HHG.

$Pr_{1-x}Ca_xMnO_3$ is a prototypical system that exhibits charge ordering. It has a pseudo-cubic perovskite structure consisting of Mn ions surrounded by oxygen octahedra. It exhibits various magnetic and electronic phases depending on the valance of the transition metal ions, temperature, and the presence of external magnetic or electric fields [26-30]. Figure 1 (a) shows the magnetic and electronic phase diagram of $Pr_{1-x}Ca_xMnO_3$ (x = 0 – 0.5) as a function of x [31]. These phases result from the couplings between spin, orbital, and charge degrees of freedom [32, 33].

We chose the hole-doping level of x = 0.4 to investigate the effect of the electronic phase on the properties of HHG. As indicated by the black arrow in the phase diagram (Fig. 1(a)), $Pr_{0.6}Ca_{0.4}MnO_3$ is a charge-disordered paramagnetic insulator (PI) at room temperature and undergoes three successive phase transitions at $T_{CO}$, $T_N$, and $T_C$ to phases that are called charge ordering (CO), antiferromagnetic (AFM), and canted-antiferromagnetic (CAF), respectively [31]. The checkerboard CO pattern in the ab plane (in the orthorhombic *Pbnm* setting) below $T_{CO} \approx 240$ K is illustrated schematically in Fig. 1(b). The transition metal cations with different valency $Mn^{3+}$ and $Mn^{4+}$ are placed alternately in the plane at low temperatures. An optical gap also opens below $T_{CO}$ (Fig. 1 (c)). The gap energy gradually increases as the temperature decreases [34]. In this system, the degeneracy of d orbital of Mn ion is also lifted due to the Jahn-Teller distortion below CO transition temperature. It has been reported that charge ordering (CO) accompanies orbital ordering (OO). As shown in the right panel of Fig. 1(b), the $e_g$-orbitals at $Mn^{3+}$ sites are ordered within the orthorhombic ab plane. This phase is denoted as CO/OO (charge-orbital ordered) phase in Fig. 1 (a). In this letter, we call this phase as CO phase.

The HHG measurements were performed with a mid-infrared (MIR) pulse laser with a photon energy $\hbar\omega$ of 0.26 eV, repetition rate of 1 kHz, pulse duration of 100 fs, and intensity of 0.3 TW/cm$^2$ at the sample surface. MIR intensity dependence of HHG yields indicates that light-matter interaction enters a non-perturbative regime in our experimental condition (see Supplemental Material [35] for detailed information on the MIR intensity dependence.) The sample temperature was controlled by a closed-cycle helium cryostat in the range of 7 K to 294 K with an accuracy of about 1 K. This allowed us to compare the HHG properties in several electronic and magnetic phases which include PM, CO, AFM, and CAF.

Figure 2 (a) shows the schematic configuration of the experimental setup in reflection geometry (details are shown in Fig. S2 of the Supplemental Material [35].) The power of the MIR light was controlled by a pair of wire-grid polarizers (WG). The polarization direction of the MIR light was controlled by a combination of a liquid crystal retarder (LCR) as a quarter-wave plate and a wire-grid polarizer. The MIR light was focused on the sample using a reflective objective lens with the 20° angle of incidence. Indium tin oxide (ITO) -coated glass was placed as a mirror that reflects the MIR light and transmits visible light.

The $Pr_{0.6}Ca_{0.4}MnO_3$ bulk crystal was mounted on the copper cold head in a cryostat with its two pseudo-cubic axes in the surface plane. Figure 2 (b) is an optical image of the sample surface (see Supplemental Material [35] for the temperature-dependent MIR reflection on the surface.) The crystal was grown using the floating zone (FZ) method. The directions of the pseudo-cubic (pc) axes corresponding to Mn-O bonds were determined with the X-ray back-reflection Laue method. In accordance with the following discussion, of the three pseudo-cubic directions, the $pc_2$ axis is assigned to the c axis, and the $pc_1$ and $pc_3$ axes are assigned perpendicular to the c axis. The surface roughness was reduced to less than 1 $\mu$m by polishing and buffing.

To study the orientation dependence of the HHG, the polarization of the MIR light was rotated to various angles ($\theta$) with respect to the $pc_1$ axis ($\theta = 0$ indicates linearly polarized MIR light along $pc_1$). Figure 2 (c) shows polar plots of the third harmonic intensity as a function of the MIR polarization angle at room temperature above $T_{CO}$ (the black plots in the center) and at the lowest temperature 7 K below $T_C$ (the blue plots). At room temperature, the high-harmonic emission is observed to be larger along the two pseudo-cubic directions, $pc_1$ and $pc_2$ parallel to Mn-O bonds, exhibiting fourfold symmetry. At 7 K, on the other hand, the high-harmonic emission is much more enhanced along $pc_1$ than along $pc_2$, exhibiting twofold symmetry (more information on the temperature dependence is in the Supplemental Material [35]).

As shown in the right panel of Fig. 1(b), below $T_{CO}$, the checkerboard CO pattern appears within the orthorhombic ab plane. One can see that along $pc_1$ and $pc_3$ axes, $Mn^{3+}$ and $Mn^{4+}$ sites alternately appear. Namely, the arrangements of charge and orbital are the same along both $pc_1$ and $pc_3$ axes. Along the orthorhombic c-axis ($pc_2$-axis), however, the same cations are aligned, which is different from the arrangements along $pc_1$ and $pc_3$ axes. Therefore, the symmetry change of the high harmonic intensity as shown in Fig. 2(c) can be attributed to the appearance of such an anisotropy owing to the CO transition. In comparison with $Ca_2RuO_4$ [22], the symmetry change of the high-harmonic intensity below the gap-opening temperature is more obvious in $Pr_{0.6}Ca_{0.4}MnO_3$. This may be due to the effect of orbital ordering, whose contribution is more prominent in perovskite-type manganese oxide [26].

Comparing the orientation dependence with the temperature-dependent anisotropic gap energy (as shown in Fig. 1(c)), we deduce that the $pc_1$ axis, along which the HHG intensity is highly temperature dependent as shown in Fig. 2(c), is in the orthorhombic ab plane and the other axis ($pc_2$) is along the orthorhombic c-axis.

Figure 2(d) shows the HH spectra measured at two temperatures below $T_C$ (7K, blue solid line) and above $T_{CO}$ (294 K, black dashed line) with the MIR laser polarized along the $pc_1$ axis. The emission of the third and the fifth harmonics were observed at 294 K, and also the seventh harmonics at 7 K. The HH intensity at 7 K is considerably larger than that at 294 K in all energy ranges of the observations. This result indicates that phase transition causes a drastic change in the HHG properties.

Detailed temperature dependences of the HH intensity are shown in Figs. 3(a) and 3(b). The HHG intensity remains at almost the same value in the PI phase above $T_{CO}$. On the other hand, it starts to gradually increase as the temperature is lowered near $T_{CO}$. No noticeable change corresponding to the magnetic phase transitions $T_N$ and $T_C$ can be seen in the intensity curve. Therefore, we conclude that the charge ordering, rather than the magnetic ordering, strongly influences HHG in $Pr_{0.6}Ca_{0.4}MnO_3$.

To reveal the relationship between HH intensity and the charge order, we plotted the high harmonic intensity as a function of the gap energy ($2\Delta$) related to the CO phase previously shown in Fig. 1(c). Figure 4(a) shows the gap energy dependence of the third harmonic intensity (filled circles) and the fifth harmonic intensity (open squares). We found that the intensity increases almost exponentially as the gap energy related to the CO phase increases. We performed the fitting of the nth order harmonic intensity with $I_n(\Delta) = I_n(0) \exp(\alpha_n \Delta)$ using an exponent of $\alpha_n$ as the fitting parameter (dashed lines in Fig. 4). The larger value of $\alpha_5 = 12.3$ eV$^{-1}$ than $\alpha_3 = 5.46$ eV$^{-1}$ indicates that fifth harmonic process is more sensitive to the gap energy than the third harmonic process. These relationships between HH intensity and the gap energy resemble those found in a previous study on the Mott insulator $Ca_2RuO_4$ [25] in which the exponent is larger for higher harmonics. Moreover, both of the exponents $\alpha_n$ ($n = 3, 5$) in Fig. 4(a) is steeper than their counterparts of the $Ca_2RuO_4$, which suggests that HHG from the CO phase in this system is more sensitive to the gap energy.

It is reasonable to suppose that thermally disturbed CO configurations (see Fig. 4(b)) in $Pr_{0.6}Ca_{0.4}MnO_3$ play a role similar to thermally disturbed antiferromagnetic spin configurations in a Mott insulator. Our results imply that the scenario proposed by Murakami *et al.* [22] applies not only to spin systems but also to CO systems without spin ordering.

Such a sensitivity of HHG to the fluctuation of local order in SCSs is no surprise. A recent study has succeeded in the reconstruction of the valence band and electron density with a spatial resolution of the order of several tens of picometers, using high harmonics [10]. Another theoretical study showed that the disorder of the atomic arrangement suppresses the HH intensity in solids [14]. These studies indicate that HHG is sensitive to the local structure with atomic-scale resolution. Our claim is consistent with these previous studies: HHG can be a new probing tool for the fluctuation of local order in

SCSs. Further comprehensive experimental and theoretical studies may lead to a deeper understanding of the physics behind the observed results, including the microscopic mechanism.

In summary, we performed HHG measurements in $Pr_{0.6}Ca_{0.4}MnO_3$ over a wide range of temperatures. We found that the charge disordering to charge ordering transition drastically changes the properties of HHG. Below $T_{CO}$, we encountered an anomalous gap energy dependence of high-harmonic intensity which resembles the one found in the Mott insulator $Ca_2RuO_4$ [25]. The resemblance between our experimental results and those reported for $Ca_2RuO_4$ suggests that a scenario similar to the one proposed by Murakami *et al.* [22] may occur in a variety of ordered phases in SCSs whether spin ordering is involved or not. We concluded that the main factor in the suppression of the high-harmonic intensity at high temperatures in our measurements is the strong coupling between the charge ordering configuration and the charge dynamics. Our results provide additional evidence for the applicability of HHG as a probing tool for the fluctuation of local order in SCSs.

**Acknowledgments**

A. N. and K. U. are thankful to Yuta Murakami for fruitful discussions. This work is supported by Grants-in-Aid for Scientific Research (S) (Grant No. JP21H05017, No. JP17H06124) and Grants-in-Aid for Scientific Research (C) (Grant No. JP22K03484). A. N. is thankful for JST SPRING, Grant No. JPMJSP2110. K. U. is thankful for a Grant-in-Aid for Challenging Research (Pioneering) (Grant No. 22K18322).

Figures & Captions

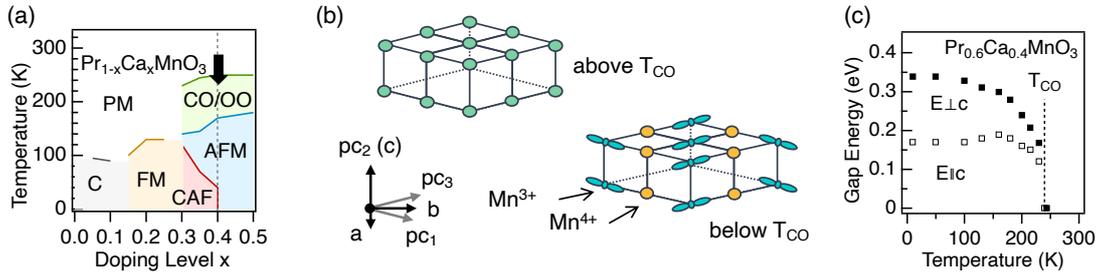

FIG. 1. Several ordered phases in $Pr_{1-x}Ca_xMnO_3$. (a) Magnetic and electronic phase diagram as a function of doping level x. PM, C, and FM denote the paramagnetic, spin-canted, and ferromagnetic phases, respectively. For $0.3 \leq x \leq 0.5$, the antiferromagnetic (AFM) insulating phase exists in the charge–orbital ordered (CO/OO) phase. The canted antiferromagnetic (CAF) phase also appears for $0.3 \leq x \leq 0.4$ (from [31]). (b) Schematic pictures of charge and orbital arrangements in the charge disordered (above $T_{CO}$) and charge ordered (below $T_{CO}$) states in $Pr_{0.6}Ca_{0.4}MnO_3$. Below $T_{CO}$, checkerboard charge ordering appears accompanied by the $e_g$-orbital ordering within the orthorhombic ab plane as shown in Fig. 3 in Ref. 26. (Arrangements of spins are omitted.) (c) Gap energy related to charge-orbital ordering as a function of temperature below $T_{CO}$ for x = 0.4 [34].

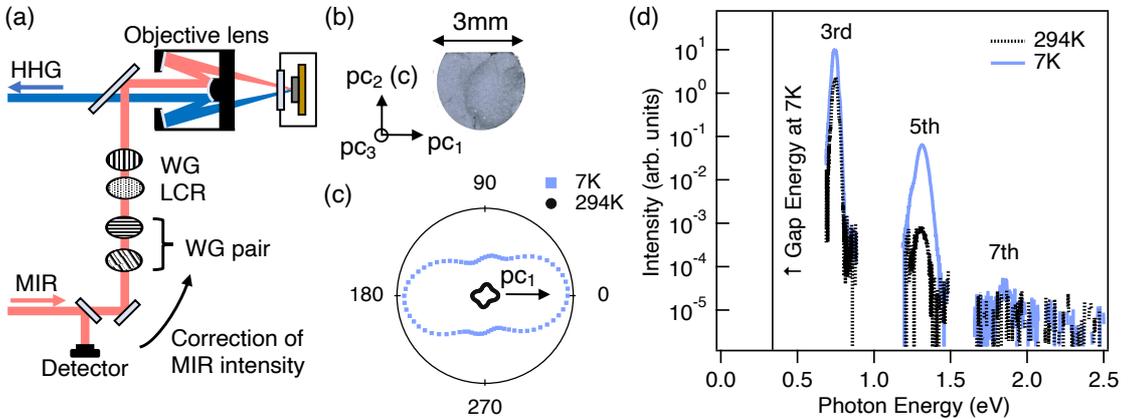

FIG. 2. High harmonic generation (HHG) in $Pr_{0.6}Ca_{0.4}MnO_3$. (a) Schematic configuration of the HHG measurement in reflection geometry. WG and LCR denote wire-grid polarizer and liquid crystal retarder, respectively. (b) Optical image of $Pr_{0.6}Ca_{0.4}MnO_3$ bulk sample. The pseudo-cubic directions (pc) corresponding to Mn-O bonds are determined by the X-ray back-reflection Laue method. (c) Polar plots of the third harmonic intensity as a function of the MIR polarization angle. The intensity increases when the MIR polarization is in the direction of the Mn-O bonds. (d) HH spectra obtained at 7 K (blue solid line) and 294 K (black dashed line). The HH spectra were obtained with the MIR laser linearly

polarized along the $pc_1$ axis.

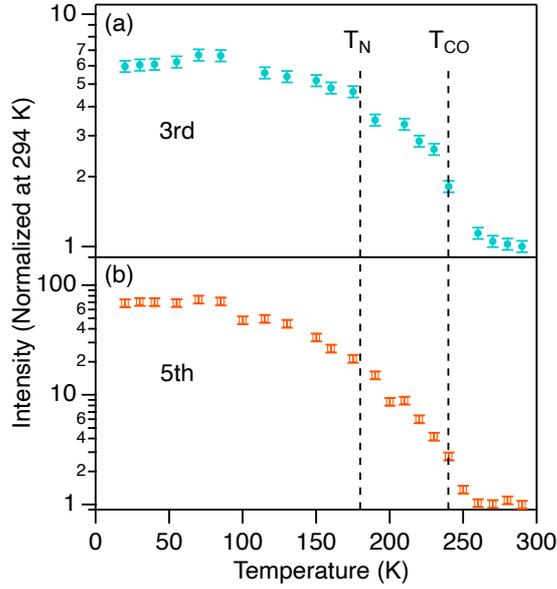

FIG. 3. Enhancement of HHG below $T_{CO}$. The temperature-dependent intensities of (a) the third, and (b) the fifth harmonics normalized at 294 K. The dashed lines at 240 K and 180 K indicate the charge-orbital ordering transition temperature ($T_{CO}$) and Néel temperature ($T_N$), respectively. Error bars are estimated by considering the fluctuations of the power of the MIR laser during the measurement.

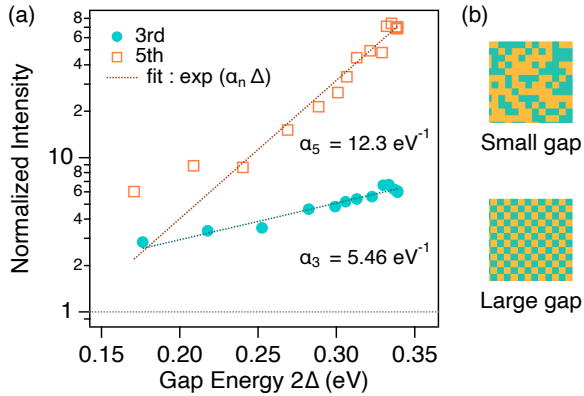

FIG. 4 (a) Gap energy dependence of HH intensity below $T_{CO}$. The intensities of the third (closed circle) and the fifth harmonics (open square) are normalized at 294 K. Here $\alpha_3$ and $\alpha_5$ are fitting parameters for each line. (b) Schematic pictures of CO configuration within the orthorhombic ab plane with small gap energy (upper panel) and large gap energy (lower panel).

# Supplemental Material: Dominant role of charge ordering on high harmonic generation in $Pr_{0.6}Ca_{0.4}MnO_3$

A. Nakano[1], K. Uchida[1], Y. Tomioka[2], M. Takaya[3], Y. Okimoto[4] and K. Tanaka[1,5]

[1] *Department of Physics, Graduate School of Science, Kyoto University, Kyoto, Kyoto 606-8502, Japan*

[2] *National Institute of Advanced Industrial Science and Technology (AIST), Tsukuba, 305-8565, Japan.*

[3] *Department of Geology and Mineralogy, Graduate School of Science, Kyoto University, Kyoto, Kyoto 606-8502, Japan*

[4] *Department of Chemistry, Tokyo Institute of Technology, Meguro, Tokyo 152-8551, Japan*

[5] *Institute for Integrated Cell-Material Sciences, Kyoto University, Sakyo-ku, Kyoto 606-8501, Japan*

## S1. Details of HHG measurements

Figure S1 shows the detailed schematic of the experimental setup. We used a regenerative amplifier with central wavelength of 800 nm, pulse duration of 35 fs, repetition rate of 1 kHz, and pulse energy of 7 mJ. The mid-infrared pulse is generated by the combination of a pair of optical parametric amplifications (OPA) and a successive difference frequency generation in GaSe. Through the DFG process, the signal outputs from two OPA with the wavelength of 1180 nm and 1566 nm are converted to CEP-stable mid-infrared pulse with wavelength of 4800 nm, and pulse duration of 100 fs. The roles of wire-grid polarizers (WG1, WG2, WG3) and a liquid crystal retarder (LCR) are described in the main text. The detecting system consists of a spectrometer and two detectors. One is a silicon-based CCD sensor detecting visible to near-infrared light and the other is an InGaAs linear sensor detecting near-infrared light.

All intensity spectra of HHG in the main text were calibrated by dividing the observed spectra by the relative detection efficiency of the system. The relative detection efficiency was determined by measuring the spectrum of a standard tungsten halogen calibration lamp. The typical accumulation time to have one spectrum was 3 seconds for the third harmonics and 30 seconds for the fifth harmonics.

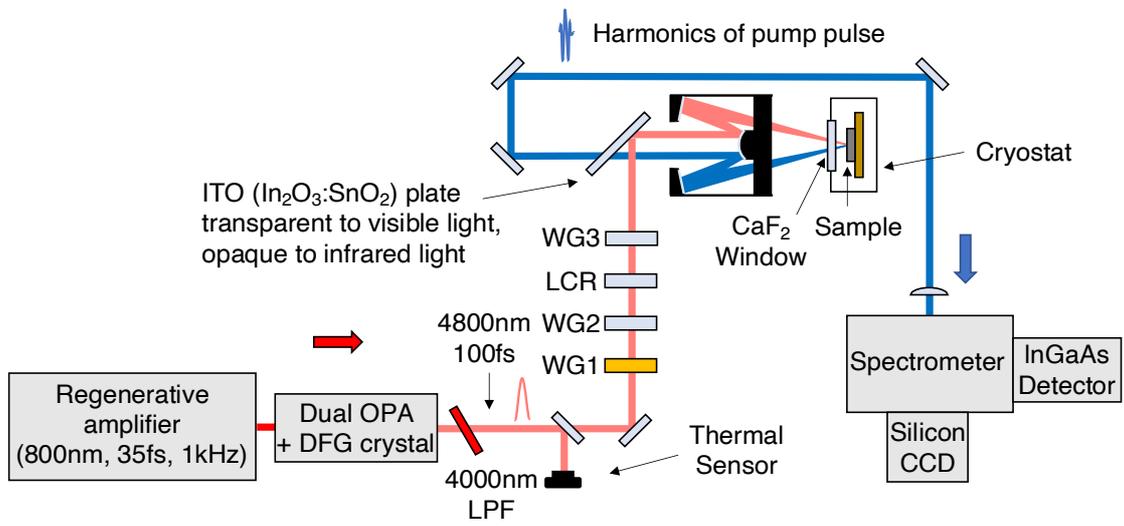

**FIG. S1**. Schematic configuration of the HHG measurement in reflection geometry. LPF denotes longpass filter.

To clarify whether the HHG measurements in the main text are in a perturbative or non-perturbative regime, we measured the intensity dependence of high-harmonics on mid-infrared (MIR) intensity by controlling the intensity of MIR incident light with a pair of wire-grid polarizers. We varied the MIR intensity at the sample surface up to 0.3 TW/cm$^2$, which corresponds to the value of the MIR intensity in Figs. 3 and 4 in the main text.

Figure S2 shows the MIR intensity dependence of the third and fifth harmonic intensities at 77 K and 295 K, which are lower and higher than the two transition temperatures, the charge order transition temperature and the Néel temperature. The dependence of the HH intensity on the MIR intensity is almost the same at the two temperatures. The third and fifth harmonic intensity is nearly proportional to the square and the cubic of the MIR intensity, respectively. This result indicates that the physical phenomenon that underlies HHG is non-perturbative light-matter interaction in our temperature range, including CO transition and magnetic transition.

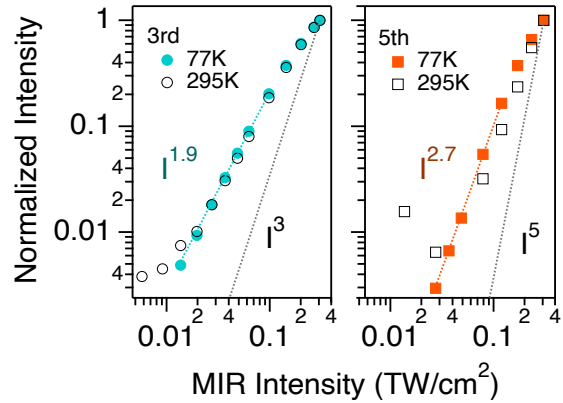

**FIG. S2.** Dependence of HH intensity on MIR intensity above and below $T_{CO}$. The third (left panel) and the fifth (right panel) harmonic intensities are plotted. The harmonic intensities are normalized by the values at the highest MIR intensity. The open markers are measured at 295 K and the solid markers are measured at 77 K.

## S2. The evaluation of Fresnel losses on the sample surface

Here we estimate the temperature-dependent Fresnel reflection loss at the sample surface to show that the increase in HH intensity at low temperatures cannot be explained solely by the suppression of reflectance of the incident light at a wavelength of 4800 nm. The roughness of the sample surface was reduced to less than 1 $\mu$m by polishing with lapping film and buffing with a soft cloth.

Figure S3 shows the values of reflectance with the MIR light polarized parallel and perpendicular to c axis. normalized at room temperature. As the temperature decreases, the amount of light not reflected at the sample surface increases. The value of reflectance only changes by 10 % over the temperature range of 80 K to 294 K. Previous research shows that the value of the reflectance at room temperature is about 0.38 [S1]. Therefore, the value of $1 - R$ changes from 0.62 to 0.66 within the temperature range.

When considering the dependence of HH intensity on MIR intensity in a previous section (Fig. S2), the third and the fifth harmonics intensity can be estimated to change by only 20 %. Therefore, the temperature dependence of the HHG intensity, which varies by one to two orders of magnitude, cannot be explained by the temperature dependence of the Fresnel reflection loss at the sample surface.

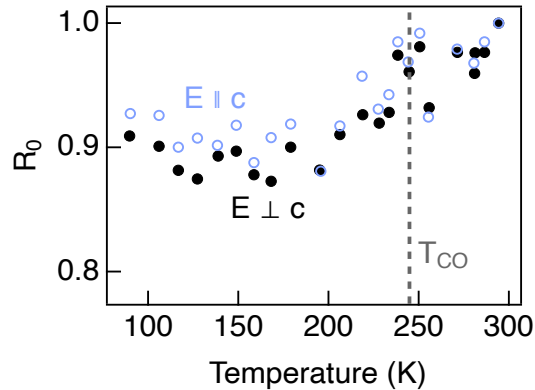

**FIG. S3.** Temperature dependences of the reflectance at the wavelength of 4800 nm normalized at room temperature. The open and filled circles correspond to the reflectance with MIR light polarized parallel and perpendicular to c axis, respectively.

## S3. Temperature-dependent anisotropy of third harmonic intensity

As shown in Fig. 2(c) in the main text, the dependence of the HHG intensity on the polarization direction of the MIR light is largely temperature dependent. Here we show the polarization direction dependence in more detail.

Figure S4(a) shows detailed orientation dependence of the intensity of the third harmonic. Each different colored line corresponds to the data obtained at different temperatures. The corresponding temperature is pointed out by colored arrows in Figure 4(b). The symmetry of the high-harmonic intensity above $T_{CO}$ is close to fourfold symmetry and is gradually reduced to twofold symmetry below $T_{CO}$. Therefore, we attributed the reduction of the symmetry to the symmetry change of the crystal accompanied by the CO transition.

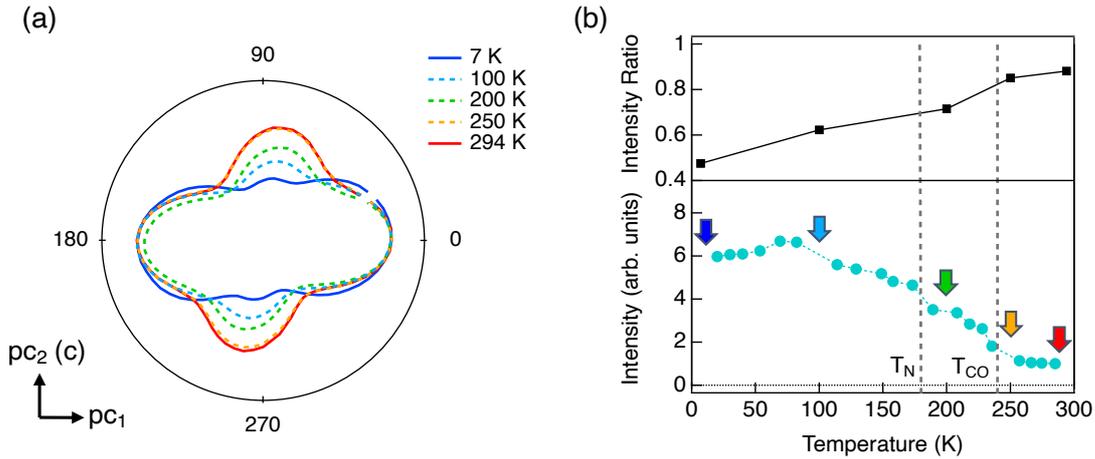

**FIG. S4.** (a) Polar plots of the third harmonic intensity as a function of the MIR polarization angle at the temperature above and below $T_{CO}$, normalized at the maximum intensity. (b) The upper panel shows the ratio of the third harmonic intensity with the MIR laser polarized along $pc_2$ axis to $pc_1$ axis. The lower panel shows linear plots of the third harmonic intensity as a function of the temperature with the MIR laser polarized along $pc_1$ axis. The color of the arrows corresponds to the color of the lines in the polar plots. The dashed lines at 240 K and 180 K indicate the charge ordering transition temperature ($T_{CO}$) and Néel temperature ($T_N$), respectively.